# The *Report [1]

Hitoshi NISHINO[2] and S. James GATES, Jr.,[3]

*Department of Physics
University of Maryland at College Park
College Park, MD 20742-4111, USA*

## Abstract

We present component and superspace formulations for the recently-proposed type IIA* (or so-called 'star') supergravity theory, which is time-like dual to the conventional type IIB theory. First, within the component approach, all terms in the action are fixed up to the quartic fermionic ones. As desired, the kinetic terms for Ramond-Ramond fields have signs opposite to the conventional case. Consistency of these are then insured by the construction of a superspace description of this theory. As a by-product, we find that a single signature parameter $s = \pm 1$ can interpolate the type IIA and type IIA* theories in superspace. This superspace result naturally allows us to present a Green-Schwarz action, that possesses $\kappa$-symmetry, consistent with such backgrounds. We also give general algebraic descriptions of such 'star' theories, so that they can be identified as representatives of some of the equivalence classes of $\kappa$-invariant Green-Schwarz actions.

---

[1]This work is supported in part by NSF grant # PHY-93-41926.
[2]E-Mail: nishino@nscpmail.physics.umd.edu
[3]E-Mail: gatess@wam.umd.edu

## 1. Introduction

Recently a new class of superstring theories called type IIA* and type IIB* theories (or so-called 'star' theories) has been presented [1]. These are related to the conventional type IIB [2] and type IIA [3][4] theories respectively by 'timelike T-dualities', which are generalization of the usual T-dualities to the time coordinates or more general spaces with indefinite signatures. The type IIA* and type IIB* theories have also been derived from M-theory by dimensional reductions on timelike circles, instead of the conventional type IIB and type IIA theories from the dimensional reductions on spacelike circles.

Even though every consideration in string-brane physics indicates that such a 'star' formulation is indeed possible [1] in ten-dimensions (10D) consistently with local supersymmetry, we still need more explicit construction of corresponding supergravity theory, including fermionic interactions which become crucial for Killing spinor equations upon compactifications. Moreover, even though the basic structure of fermionic interaction terms can be easily conjectured, it is still advantageous to clarify these interaction terms explicitly, excluding their sign/factor ambiguity.

In this paper, we present an explicit formulation of type IIA* supergravity theory, namely we give a lagrangian for an invariant action up to quartic fermionic terms, and supersymmetry transformation rules. Correspondingly, we give an equivalent superspace formulation which acts as the foundation for a Green-Schwarz superstring formulation in the presence of such a supergravity background. This also provides a confirmation for the existence of the yet-to-be determined quartic fermion terms, and allows the complete determination of the component theory. We first used this type of approach many years ago [5], to ascertain whether the incomplete proposal of the massive type IIA theory [6] would admit quartic fermion terms.

## 2. Component Formulation

We start with our result for the component formulation of type IIA* theory [1]. The field content for the type IIA* theory [1] looks formally the same as that of the type IIA theory [3][4], *i.e.*, $(e_\mu{}^m, \psi_\mu, A_\mu, B_{\mu\nu}, C_{\mu\nu\rho}, \varphi, \chi)$, where the potential fields $A_\mu$, $B_{\mu\nu}$ and $C_{\mu\nu\rho}$ respectively have the second, third and fourth-rank field strengths: $F_{\mu\nu}$, $G_{\mu\nu\rho}$, $H_{\mu\nu\rho\sigma}$, while the gravitino $\psi_\mu$ and the dilatino $\chi$ are both Majorana, reflecting the non-chiral feature of the type IIA* theory.

Our lagrangian for an invariant action is given by[4]

$$e^{-1}\mathcal{L} = -\tfrac{1}{4}R - \tfrac{i}{2}(\overline{\psi}_\mu\gamma_{11}\gamma^{\mu\nu\rho}D_\nu\psi_\rho) + \tfrac{1}{12}e^{2\varphi}G_{[3]}{}^2 + \tfrac{1}{2}(\partial_\mu\varphi)^2$$
$$- \tfrac{i}{2}(\overline{\chi}\gamma_{11}\gamma^\mu D_\mu\chi) + \tfrac{1}{48}e^{-\varphi}H'_{[4]}{}^2 + \tfrac{1}{4}e^{-3\varphi}F_{[2]}{}^2$$
$$+ \tfrac{1}{\sqrt{2}}(\overline{\chi}\gamma_{11}\gamma^\mu\gamma^\nu\psi_\mu)\partial_\nu\varphi + \tfrac{1}{1152}e^{-1}\epsilon^{[4][4]'[2]}H_{[4]}H_{[4]'}B_{[2]}$$
$$- \tfrac{1}{96}e^{-\varphi/2}\,[\,(\overline{\psi}_\lambda\gamma^{\lambda\omega\mu\nu\rho\sigma}\psi_\omega) + 12(\overline{\psi}^\mu\gamma^{\nu\rho}\psi^\sigma) + \tfrac{i}{\sqrt{2}}(\overline{\chi}\gamma^\tau\gamma^{\mu\nu\rho\sigma}\psi_\tau) - \tfrac{3}{4}(\overline{\chi}\gamma^{\mu\nu\rho\sigma}\chi)\,]\,H'_{\mu\nu\rho\sigma}$$
$$+ \tfrac{1}{24}e^\varphi\,[\,i(\overline{\psi}_\sigma\gamma^{\sigma\tau\mu\nu\rho}\psi_\tau) - 6i(\overline{\psi}^\mu\gamma^\nu\psi^\rho) + \sqrt{2}(\overline{\chi}\gamma^\tau\gamma^{\mu\nu\rho}\psi_\tau)\,]\,G_{\mu\nu\rho}$$
$$+ \tfrac{1}{8}e^{-3\varphi/2}\,[\,(\overline{\psi}_\rho\gamma_{11}\gamma^{\rho\sigma\mu\nu}\psi_\sigma) + 2(\overline{\psi}^\mu\gamma_{11}\psi^\nu) - \tfrac{3i}{\sqrt{2}}(\overline{\chi}\gamma_{11}\gamma^\rho\gamma^{\mu\nu}\psi_\rho) - \tfrac{5}{4}(\overline{\chi}\gamma_{11}\gamma^{\mu\nu}\chi)\,]\,F_{\mu\nu} \ . \quad (2.1)$$

Our notation is essentially the same as in [3], *except* that our $\gamma_{11}$-matrix is defined by $\gamma_{11} \equiv (1/10!)\epsilon^{m_1\cdots m_{10}}\gamma_{m_1\cdots m_{10}} \equiv \gamma_0\gamma_1\cdots\gamma_9$, so that $(\gamma_{11})^2 \equiv +I$. Relevantly, we also adopt the index conventions, such as $_{[3]}$ for totally antisymmetric indices to save space, *e.g.*, $A_{[3]}B^{[3]} \equiv A_{\mu\nu\rho}B^{\mu\nu\rho}$. The supersymmetry transformation rules are

$$\delta_Q e_\mu{}^m = -i(\overline{\epsilon}\gamma_{11}\gamma^m\psi_\mu) \ , \quad \delta_Q\varphi = +\tfrac{1}{\sqrt{2}}(\overline{\epsilon}\gamma_{11}\chi) \ ,$$
$$\delta_Q\psi_\mu = D(\widehat{\omega})\epsilon + \tfrac{i}{32}e^{-3\varphi/2}(\gamma_\mu{}^{\rho\sigma} - 14\delta_\mu{}^\rho\gamma^\sigma)\epsilon\,\widehat{F}_{\rho\sigma} + \tfrac{1}{48}e^\varphi\gamma_{11}(\gamma_\mu{}^{\nu[2]} - 9\delta_\mu{}^\nu\gamma^{[2]})\epsilon\,\widehat{G}_{\nu[2]}$$
$$+ \tfrac{i}{128}e^{-\varphi/2}\gamma_{11}(\gamma_\mu{}^{\nu[3]} - \tfrac{20}{3}\delta_\mu{}^\nu\gamma^{[3]})\epsilon\,\widehat{H}'_{\nu[3]} \ ,$$
$$\delta_Q\chi = -\tfrac{i}{\sqrt{2}}\gamma^\mu\epsilon\widehat{D}_\mu\varphi - \tfrac{3}{8\sqrt{2}}e^{-3\varphi/2}\gamma^{[2]}\epsilon\,\widehat{F}_{[2]} + \tfrac{i}{12\sqrt{2}}e^\varphi\gamma_{11}\gamma^{[3]}\epsilon\,\widehat{G}_{[3]} + \tfrac{1}{96\sqrt{2}}e^{-\varphi/2}\gamma_{11}\gamma^{[4]}\epsilon\,\widehat{H}'_{[4]} \ ,$$
$$\delta_Q A_\mu = -\tfrac{1}{2}e^{3\varphi/2}(\overline{\epsilon}\gamma_{11}\psi_\mu) - \tfrac{3i}{4\sqrt{2}}e^{3\varphi/2}(\overline{\epsilon}\gamma_{11}\gamma_\mu\chi) \ ,$$
$$\delta_Q B_{\mu\nu} = +ie^{-\varphi}(\overline{\epsilon}\gamma_{[\mu}\psi_{\nu]}) + \tfrac{1}{2\sqrt{2}}e^{-\varphi}(\overline{\epsilon}\gamma_{\mu\nu}\chi) \ ,$$
$$\delta_Q C_{\mu\nu\rho} = +\tfrac{3}{2}e^{\varphi/2}(\overline{\epsilon}\gamma_{[\mu\nu}\psi_{\rho]}) + \tfrac{i}{4\sqrt{2}}e^{\varphi/2}(\overline{\epsilon}\gamma_{\mu\nu\rho}\chi) + 6A_{[\mu}(\delta_Q B_{\nu\rho]}) \ . \quad (2.2)$$

The superfield strengths are defined, as in [3], as,

$$\widehat{F}_{\mu\nu} \equiv 2\partial_{[\mu}A_{\nu]} + \tfrac{1}{2}e^{3\varphi/2}(\overline{\psi}_\mu\gamma_{11}\psi_\nu) + \tfrac{3i}{2\sqrt{2}}e^{3\varphi/2}(\overline{\psi}_{[\mu|}\gamma_{11}\gamma_{|\nu]}\chi) \ ,$$
$$\widehat{G}_{\mu\nu\rho} \equiv 3\partial_{[\mu}B_{\nu\rho]} - \tfrac{3i}{2}e^{-\varphi}(\overline{\psi}_{[\mu}\gamma_\nu\psi_{\rho]}) - \tfrac{3}{2\sqrt{2}}e^{-\varphi}(\overline{\psi}_{[\mu}\gamma_{\nu\rho]}\chi) \ ,$$
$$\widehat{H}'_{\mu\nu\rho\sigma} \equiv 4\partial_{[\mu}C_{\nu\rho\sigma]} + 8A_{[\mu}G_{\nu\rho\sigma]} - 3e^{\varphi/2}(\overline{\psi}_{[\mu}\gamma_{\nu\rho}\psi_{\sigma]}) - \tfrac{i}{\sqrt{2}}e^{\varphi/2}(\overline{\psi}_{[\mu}\gamma_{\nu\rho\sigma]}\chi)$$
$$\equiv \widehat{H}_{\mu\nu\rho\sigma} + 8A_{[\mu}G_{\nu\rho\sigma]} \ . \quad (2.3)$$

In this section of component formulation, the antisymmetrization symbol $_{[\mu\nu]}$, *etc.* is normalized, *e.g.*, $P_{[\mu}Q_{\nu]} \equiv (1/2)(P_\mu Q_\nu - P_\nu Q_\mu)$.

There are some remarks in order: First, the kinetic terms for the Ramond-Ramond (RR) bosonic fields have 'wrong' sign, as expected from [1]. Second, we found the kinetic terms for fermionic fields require factors of the $\gamma_{11}$-matrix to appear. This is reflected by the fact that our supersymmetry is dictated by the algebra

$$\{Q_{\underline{\alpha}}{}^{\pm}, Q_{\underline{\beta}}{}^{\mp}\} = \pm(\mathcal{P}^{\pm}\gamma^m\mathcal{P}^{\mp})_{\underline{\alpha}\underline{\beta}} P_m \ , \qquad (2.4)$$

where $\mathcal{P}^{\pm} \equiv (I \pm \gamma_{11})/2$ are the usual chiral projection operators for the indices $^{\pm}$. Third, we found that the patterns of the appearance of the $\gamma_{11}$-matrix in various terms in the system is much like the replacements of $\gamma^m \to \gamma_{11}\gamma^m$, in addition to $(A_\mu, C_{\mu\nu\rho}) \to (-iA_\mu, -iC_{\mu\nu\rho})$ in [1]. Even though there arise some subtle sign flips at many places, we found that what is happening can be easily understood universally by these replacements. This allows the identification of a transformation that acts on the gamma matrices together with the RR sector fields as the origin of the type IIA$^*$ theory relative to the standard type IIA theory. Fourth, we found that various signs for other terms, such as that in the Chern-Simons term, or those in the exponents proportional to the dilaton $\varphi$ stay the same as in the type IIA$^*$ case [3][4]. Fifth, we found that the exponential dependence on the dilaton $\varphi$ appearing in the lagrangian is exactly the same as in the type IIA theory, despite of the 'wrong' signs for the kinetic terms for the RR fields. This will result in subtle sign flips for RR field strength terms in the dilaton field equations, that may potentially cause desirable possibilities as well as obstructions for compactifications. Relevantly, our lagrangian has the global scale invariance for the constant dilaton shift:

$$\varphi \to \varphi + c \ , \quad A_\mu \to e^{3c/2} A_\mu \ , \quad B_{\mu\nu} \to e^{-c} B_{\mu\nu} \ , \quad C_{\mu\nu\rho} \to e^{c/2} C_{\mu\nu\rho} \ . \qquad (2.5)$$

---

[4] As usual, the Latin (or Greek) indices in component formulation are for the local (or curved) coordinates. Our signature is $(\eta_{mn}) = \text{diag.}\,(+, -, \cdots, -)$.

Our result does not contradict the common wisdom [7] that the dilaton field decouples from the RR fields. This is because all of our RR fields are canonical, making such dilaton dependence unavoidable.[5] Sixth, we can also interpret this type IIA$^*$ theory, in terms of $N = 1$ multiplets. Namely the conventional type IIA multiplet is composed of two independent $N = 1$ multiplets of supergravity (SG) $(e_\mu{}^m, \psi_\mu{}^+, \chi^-, B_{\mu\nu}, \varphi)$, and the $N = 1$ matter tensor multiplet (TM) $(\psi_\mu{}^-, C_{\mu\nu\rho}, \chi^+, A_\mu)$. Note that we can have not only the conventional total action $I_{\text{SG}} + I_{\text{TM}}$, but also an alternative action $I_{\text{SG}} - I_{\text{TM}}$. In particular, $\gamma_{11}$ in the kinetic terms of the fermionic fields in (2.1) in the 32-component notation is consistent with supersymmetric invariance. In terms of superstring language, the bosonic fields in the multiplet SG correspond to the Neveu-Schwarz (NS) sector, and those in the TM to the RR sector for bosonic fields. This is why we can flip the overall sign of the 'matter' lagrangian $I_{\text{TM}}$. It also seems that such freedom of flipping signs only for some the 'matter' lagrangians is common to all supergravity theories even in lower dimensions.

---

[5]This feature will be more elucidated in superspace, when dealing with what is called the set of '$\beta$-function favored constraints'.

The subtle sign changes for the RR sector are also reflected in the bosonic field equations:

$$R_{\mu\nu} = +\tfrac{1}{96}e^{-\varphi}(32H'_{\mu[3]}H'_\nu{}^{[3]} - 3g_{\mu\nu}H'_{[4]}{}^2) + \tfrac{1}{12}e^{2\varphi}(12G_{\mu[2]}G_\nu{}^{[2]} - g_{\mu\nu}G_{[3]}{}^2)$$
$$+ \tfrac{1}{8}e^{-3\varphi}(16F_{\mu\rho}F_\nu{}^\rho - g_{\mu\nu}F_{[2]}{}^2) + 2(\partial_\mu\varphi)(\partial_\nu\varphi) ~, \tag{2.6}$$

$$\partial_\nu(ee^{-3\varphi}F^{\mu\nu}) + \tfrac{1}{3}e^{-\varphi}H'^{\mu[3]}G_{[3]} = 0 ~, \tag{2.7}$$

$$\partial_\rho(ee^{2\varphi}G^{\mu\nu\rho} - 2ee^{-\varphi}H'^{\mu\nu\rho\sigma}A_\sigma) + \tfrac{1}{576}\epsilon^{\mu\nu[4][4]'}H_{[4]}H_{[4]'} = 0 ~, \tag{2.8}$$

$$\partial_\sigma(ee^{-\varphi}H'^{\mu\nu\rho\sigma}) + \tfrac{1}{72}\epsilon^{\mu\nu\rho[4][3]}H_{[4]}G_{[3]} = 0 ~, \tag{2.9}$$

$$D_\mu^2\varphi + \tfrac{1}{48}e^{-\varphi}H'_{[4]}{}^2 - \tfrac{1}{6}e^{2\varphi}G_{[3]}{}^2 + \tfrac{3}{4}e^{-3\varphi}F_{[2]}{}^2 = 0 ~. \tag{2.10}$$

The 'wrong' signs for the kinetic terms of the RR fields are now reflected in the relative sign between the $H^2, F^2$ and $G^2$-terms in (2.10). In the conventional type IIA theory [3][4], this relative sign is positive, and therefore $G_{[3]}$ can develop non-trivial background for spatial directions. In the present case of type IIA$^*$ [1][8], due to the flipped relative sign for these three terms in (2.10), $G_{[3]}$ can develop non-trivial background containing the time coordinate, in order for a similar cancellation between the three terms to take place, within our signature convention $(+, -, \cdots, -)$. As described in [8], this can be understood such as the $U(1)$ fibrations in the timelike direction over the non-compact manifold $\widetilde{CP}^2$.

## 3. Superspace Formulation

Once the component formulation has been established, the corresponding superspace formulation is rather straightforward. Here we list our constraint set for future reference. Our superfield strengths are $F_{AB}, G_{ABC}, H_{ABCD}, T_{AB}{}^C, R_{AB}{}^{cd}$ in self-explanatory forms, satisfying the same basic Bianchi identities as the type IIA case [5][9][6][10], or more explicitly,

$$\tfrac{1}{2}\nabla_{[A}F_{BC)} - \tfrac{1}{2}T_{[AB|}{}^D F_{D|C)} \equiv 0 ~, \tag{3.1}$$

$$\tfrac{1}{6}\nabla_{[A}G_{BCD)} - \tfrac{1}{4}T_{[AB|}{}^E G_{E|CD)} \equiv 0 ~, \tag{3.2}$$

$$\tfrac{1}{24}\nabla_{[A}H_{BCDE)} - \tfrac{1}{12}T_{[AB|}{}^F H_{F|CDE)} - \tfrac{1}{12}F_{[AB}G_{CDE)} \equiv 0 ~, \tag{3.3}$$

$$\tfrac{1}{2}\nabla_{[A}T_{BC)}{}^D - \tfrac{1}{2}T_{[AB|}{}^E T_{E|C)}{}^D - \tfrac{1}{4}R_{[AB|e}{}^f(\mathcal{M}_f{}^e)_{|C)}{}^D \equiv 0 ~. \tag{3.4}$$

In this section of superspace, our (anti)symmetrization is defined by $P_{[A}Q_{B)} \equiv P_A Q_B \mp P_B Q_A$, with *no* normalization. Reflecting our component result, our $H$-Bianchi identity (3.3) has the same factor and sign as the type IIA case [10] for its Chern-Simons term. As usual in superspace, we can fix our constraints, satisfying all the Bianchi identities up to dimension $d \leq 1$. In principle, there are infinitely many equivalent sets of superspace constraints, which are related to each other *via* super Weyl rescalings [11]. However, there is the simplest set called '$\beta$-function favored constraint' ($\beta$FFC), first introduced in order to simplify the $\beta$-function computation in the Green-Schwarz formulation [12], and used also in [9] for the type IIA theory. Even though the usage of the $\beta$FFC has some drawbacks, when comparing superspace result with 'canonical' component ones, we adopt this set in this paper due to its simplicity. We should also mention that $\beta$FFC constraints correspond to the use of the so-called 'string-frame' formulation of the component theories even though the discovery of $\beta$FFC constraints preceded the latter by some time.

---

[6]Some coefficients for the type IIA superspace constraints in [9] were corrected in [10].

In order to compare the constraints for the type IIA* with the conventional type IIA theory, we use a convenient signature parameter $s = \pm 1$, which switches from the former to the latter. This comparison can be most easily done by studying the bosonic field equations, as will be seen later. Our set of $\beta$FFC constraints for type IIA* or type IIA theory is now summarized as

$$\begin{aligned}
&T_{\alpha\beta}{}^c = +i(\sigma^c)_{\alpha\beta} \ , \qquad T_{\dot\alpha\dot\beta}{}^c = -is(\sigma^c)_{\dot\alpha\dot\beta} \ , \\
&T_{\alpha\beta}{}^\gamma = +\delta_{(\alpha}{}^\gamma \chi_{\beta)} + (\sigma^c)_{\alpha\beta}(\sigma_c\chi)^\gamma \ , \qquad T_{\dot\alpha\dot\beta}{}^{\dot\gamma} = +\delta_{(\dot\alpha}{}^{\dot\gamma} \chi_{\dot\beta)} + (\sigma^c)_{\dot\alpha\dot\beta}(\sigma_c\chi)^{\dot\gamma} \ , \\
&T_{ab}{}^\gamma = -\tfrac{1}{8}(\sigma^{cd})_\alpha{}^\gamma G_{bcd} \ , \qquad T_{\dot\alpha b}{}^{\dot\gamma} = +\tfrac{1}{8}(\sigma^{cd})_{\dot\alpha}{}^{\dot\gamma} G_{bcd} \ , \\
&T_{\alpha b}{}^{\dot\gamma} = +\tfrac{i}{16}(\sigma_b\sigma^{cd})_\alpha{}^{\dot\gamma} e^{-\varphi} F_{cd} - \tfrac{i}{192}s(\sigma_b\sigma^{[4]})_\alpha{}^{\dot\gamma} e^{-\varphi} H_{[4]} \\
&\qquad + \tfrac{i}{8}s(\sigma_b)_\alpha{}^{\dot\gamma}(\chi\chi) - \tfrac{i}{16}s(\sigma_b\sigma^{cd})_\alpha{}^{\dot\gamma} \chi_{cd} + \tfrac{i}{192}s(\sigma_b\sigma^{[4]})_\alpha{}^{\dot\gamma} \chi_{[4]} \ , \\
&T_{\dot\alpha b}{}^\gamma = +\tfrac{i}{16}s(\sigma_b\sigma^{cd})_{\dot\alpha}{}^\gamma e^{-\varphi} F_{cd} + \tfrac{i}{192}(\sigma_b\sigma^{[4]})_{\dot\alpha}{}^\gamma e^{-\varphi} H_{[4]} \\
&\qquad - \tfrac{i}{8}(\sigma_b)_{\dot\alpha}{}^\gamma(\chi\chi) - \tfrac{i}{16}(\sigma_b\sigma^{cd})_{\dot\alpha}{}^\gamma \chi_{cd} - \tfrac{i}{192}(\sigma_b\sigma^{[4]})_{\dot\alpha}{}^\gamma \chi_{[4]} \ , \\
&F_{\alpha\dot\beta} = +C_{\alpha\dot\beta} e^\varphi \ , \qquad F_{\alpha b} = -ise^\varphi(\sigma_b\chi)_\alpha \ , \qquad F_{\dot\alpha b} = -ie^\varphi(\sigma_b\chi)_{\dot\alpha} \ , \\
&G_{\alpha\beta c} = +i(\sigma^c)_{\alpha\beta} \ , \qquad G_{\dot\alpha\dot\beta c} = +is(\sigma^c)_{\dot\alpha\dot\beta} \ , \\
&H_{\alpha\dot\beta cd} = +e^\varphi(\sigma_{cd})_{\alpha\dot\beta} + Y_{\alpha\dot\beta cd} \ , \\
&H_{\alpha bcd} = -ise^\varphi(\sigma_{bcd}\chi)_\alpha + Y_{\alpha bcd} \ , \qquad H_{\dot\alpha bcd} = +ie^\varphi(\sigma_{bcd}\chi)_{\dot\alpha} + Y_{\dot\alpha bcd} \ , \\
&\nabla_\alpha \varphi = +\chi_\alpha \ , \qquad \nabla_{\dot\alpha}\varphi = +\chi_{\dot\alpha} \ , \qquad \nabla_\alpha \chi_{\dot\beta} = -\nabla_{\dot\beta}\chi_\alpha \ , \\
&\nabla_\alpha \chi_\beta = +\tfrac{i}{2}(\sigma^c)_{\alpha\beta}\nabla_c\varphi + \tfrac{i}{24}s(\sigma^{[3]})_{\alpha\beta} G_{[3]} - \chi_\alpha \chi_\beta \ , \\
&\nabla_{\dot\alpha}\chi_{\dot\beta} = -\tfrac{i}{2}s(\sigma^c)_{\dot\alpha\dot\beta}\nabla_c\varphi + \tfrac{i}{24}(\sigma^{[3]})_{\dot\alpha\dot\beta} G_{[3]} - \chi_{\dot\alpha}\chi_{\dot\beta} \ , \\
&\nabla_\alpha \chi_{\dot\beta} = +\tfrac{3}{16}s(\sigma^{cd})_{\alpha\dot\beta} e^{-\varphi} F_{cd} - \tfrac{1}{192}(\sigma^{[4]})_{\alpha\dot\beta} e^{-\varphi} H_{[4]} \\
&\qquad + \tfrac{5}{8} C_{\alpha\dot\beta}(\chi\chi) - \tfrac{3}{16}(\sigma^{[2]})_{\alpha\dot\beta}\chi_{[2]} + \tfrac{1}{192}(\sigma^{[4]})_{\alpha\dot\beta}\chi_{[4]} \ . 
\end{aligned} \qquad (3.5)$$

Here $\chi_{[2n]} \equiv \chi^\alpha(\sigma_{[2n]})_\alpha{}^\beta \chi_\beta$, and $(\chi\chi) \equiv \chi^\alpha \chi_\alpha$. As is also explicit from these forms, we are using the 16 component Majorana-Weyl chiral spinor indices, *i.e.,* the undotted ones $\alpha, \beta, \cdots = 1, 2, \cdots, 16$ are for the positive chirality, while the dotted ones $\dot\alpha, \dot\beta, \cdots = \dot1, \dot2, \cdots, \dot{16}$ for the negative chirality. Due to this chiral notation, we use here the $\sigma$-matrices instead of the $\gamma$-matrices in the component results. As usual in superspace, other independent components, such as $G_{\alpha bc}$ are zero. The $Y_{ABCD}$ is the super Chern-Simons form defined by [5]

$$Y_{ABCD} \equiv \tfrac{1}{4} F_{[AB} B_{CD)} \ . \tag{3.6}$$

Let us once more emphasize that (3.5) is a consistent superspace description of type IIA$^*$ as well as type IIA theory, by switching the signature of the parameter $s = \pm 1$. The consistency of this superspace description implies a component formulation exists complete with quartic fermion terms its action. Such a component theory is related *via* a set of field redefinitions to the canonical component theory discussed in Section two, and thus insures that the system also possesses a unique and well defined set of quartic fermion terms for its complete description.

Most of the results in (3.5) are similar to those in [9][10], except for several sign flips reflecting the involvement of $\gamma_{11}$ in the 32 component spinor notation in the previous section. In particular, our non-standard supersymmetry algebra for type IIA$^*$ involving the $\gamma_{11}$ as in (2.3), is reflected in the sign difference of $T_{\dot\alpha\dot\beta}{}^c$ from $T_{\alpha\beta}{}^c$, depending on $s = \pm 1$. The same is also true for $G_{\dot\alpha\dot\beta c}$.

We now come to the point to see which case out of $s = +1$ and $s = -1$ corresponds to the type IIA$^*$ theory. To see this most effectively, we look into the scalar curvature superfield equation. To this end, we first derive the gravitino/dilatino superfield equation, taking the usual step out of the $T$-Bianchi identity of the $(\alpha\beta c, \delta)$-type [13][12]. It turns out to that the gravitino/dilatino superfield equation is independent of the value of $s = \pm 1$, as

$$i(\sigma^b T_{ab})_\alpha + 2\nabla_a \chi_\alpha - \tfrac{1}{4}(\sigma^{bc}\chi)_\alpha G_{abc} = 0 \ . \tag{3.7}$$

Now the scalar curvature superfield equation can be obtained as usual by applying another spinorial derivative on (3.7), with one $\sigma$-matrix contracting all the spinorial indices. Ignoring fermionic bilinear terms, it turns out to be[7]

$$R + \tfrac{1}{4} G_{[3]}{}^2 - \tfrac{3}{4} s e^{-2\varphi} F_{[2]}{}^2 - \tfrac{1}{48} s e^{-2\varphi} H_{[4]}{}^2 + 2\nabla_a^2 \varphi = 0 \ , \tag{3.8}$$

On the other hand, from our component result, the trace of the gravitational field equation (2.6) should result in the same relative signatures for all the $F^2$, $G^2$ and $H^2$-terms for

---

[7]The absence of the $(\nabla_a \varphi)^2$-term here is natural, because it is similar to the type I supergravity in [12].

the type IIA* theory.[8] In other words, in (3.8), the values of $s$ are

$$s = \begin{cases} +1 & \text{(for type IIA theory)}, \\ -1 & \text{(for type IIA* theory)}. \end{cases} \quad (3.9)$$

Considering this point, we find that the papers [5][9] or [10] with the choice of $s = +1$ actually give the type IIA* theory, instead of the conventional type IIA theory.[9] It is thus amusing that just one signature parameter $s = \pm 1$ can interpolate between the type IIA and type IIA* $\beta$FFC in such a compact form.

Careful readers may wonder, if the component result (2.2) contradict (3.9), according to the usual rule of getting component transformation rule from the superspace constraints, as described in page 323 of ref. [14]. However, this can be easily understood by the ambiguity of assigning either $+\psi_m{}^{\dot{\alpha}}$ or $-\psi_m{}^{\dot{\alpha}}$ for the negative chirality components for the gravitino, *i.e.*, the relatively positive or negative sign for the dotted component in $\psi_m{}^{\underline{\alpha}}$ compared with its undotted component $\psi_m{}^\alpha$.[10] Thus, in terms of superspace language, the ambiguity between type IIA and type IIA* theories seem to arise from this signature assignment ambiguity.

An important lesson has emerged from the present result, *i.e.*, when solving superspace Bianchi identities, one must confirm also the field equations in order to distinguish between the conventional and 'star' theories. In other words, a consistent solutions to superspace Bianchi identities is *not* unique.

According to our gravitational field equation (3.8), the dilaton $\varphi$ seem to couple to the RR-field strengths $F$ and $H$, while not to the $G$-field strength. One may wonder if this contradicts the common wisdom that the dilaton does not couple to the RR-fields. However, this can be easily understood from the fact that these exponential couplings in eq. (3.8) can be deleted by an overall exponential factor $e^{+2\varphi}$, *e.g.*, the first term becomes $e^{2\varphi}R$. In fact, we expect that the $\beta$FFC system will have the Hilbert action with such a Brans-Dicke type dependence on the dilaton.

In this paper we started with the component formulation, and subsequently we gave the superspace formulation for the $\beta$FFC. The reason for this is that a superspace formulation, in particular the $\beta$FFC, is more difficult than component formulations for getting an invariant lagrangian, which is crucial for our purpose of confirming subtle sign flips for the RR kinetic terms. It is the component formulation that can first guide us to the right lagrangian with the transformation rules, despite of the usual difficulty with handling purely fermionic terms. On the other hand, we have also seen that the superspace formulation is easier to unify two theories of type IIA and type IIA$^*$, in terms of just one signature parameter interpolating the two $\beta$FFC sets. The similarity between these systems also indicate their close relationship under some duality like T-duality, as has been suggested in [1].

---

[8]These relative signs are not supposed to be changed by any super Weyl rescalings [11] such as from the canonical set to the $\beta$FFC.

[9]This is not surprising, because in the papers [5][9][10], no field equations had been derived in order to confirm the signs of kinetic terms. It is only after we have derived the field equations that we can see whether the system corresponds to the type IIA or type IIA* theory.

[10]Note that this is also related to the fact that in 10D, the dotted and undotted spinors are *not* related by complex conjugation, *unlike* the 4D case.

## 4. Green-Schwarz Action

Based on our superspace formulation, we can look at the corresponding Green-Schwarz formulation [7] for the type IIA$^*$ superstring theory. This turns out to be rather straightforward, because of the parallel structures between the type IIA and type IIA$^*$ superspaces.

As we have seen, the type IIA$^*$ system has superspace constraints with sign flips only in limited terms, such as $G_{\dot\alpha\dot\beta c}$ compared with $G_{\alpha\beta c}$. Similar ambiguities had been already pointed out in 4D context in [15][16], *i.e.*, the $\kappa$-invariance of the action allows sign ambiguity in the components $G_{\underline{\alpha\beta} c}$.[11]

We start with the action for Green-Schwarz superstring on the 10D type IIA and type IIA$^*$ backgrounds [17]:

$$I \equiv \int d^2\xi\, V^{-1}\left(\eta_{ab}\Pi_+{}^a\Pi_-{}^b + \Pi_+{}^A\Pi_-{}^B B_{BA}\right) \quad, \tag{4.1}$$

where $V \equiv \det(V_\pm{}^i)$ is the determinant of the zweibein $V_\pm{}^i$, and the indices $i, j, \cdots = 0, 1$ are for the curved 2D coordinates $\xi^i$, while $\pm$ are for the local Lorentz light-cone coordinates. As usual, $\Pi_\pm{}^A$ is the pull-back: $\Pi_\pm{}^A \equiv (\partial_\pm Z^M) E_M{}^A$ with the superspace coordintates $Z^M$ and the (inverse)vielbein $E_M{}^A$. At this point, we do not specify the choice between the type IIA and type IIA$^*$ backgrounds, for the reason to be clarified shortly.

The total action (4.1) is invariant under the following $\kappa$-symmetry:

$$\delta_\kappa V_+{}^i = +2(\Pi_+{}^\alpha \kappa_{+\alpha}) V_-{}^i \quad, \qquad \delta_\kappa V_-{}^i = +2(\Pi_-{}^{\dot\alpha} \kappa_{-\dot\alpha}) V_+{}^i \quad, \tag{4.2a}$$

$$\delta_\kappa E^\alpha = +i\,(\slashed{\Pi}_-)^{\alpha\beta} \kappa_{+\beta} \quad, \qquad \delta_\kappa E^{\dot\alpha} = -is\,(\slashed{\Pi}_+)^{\dot\alpha\dot\beta} \kappa_{-\dot\beta} \quad, \tag{4.2b}$$

$$\delta_\kappa(V^{-1}) = 0 \quad, \qquad \delta_\kappa E^a = 0 \quad, \tag{4.2c}$$

where $(\slashed{\Pi}_-)^{\underline{\alpha\beta}} \equiv (\sigma_c)^{\underline{\alpha\beta}} \Pi_-{}^c$, and $\delta_\kappa E^A \equiv (\delta_\kappa Z^M) E_M{}^A$. The constant $s$ in (4.2b) is the same as in (3.5), and the invariance of $I$ under (4.2) is valid for both values of $s = \pm 1$. Therefore, (4.1) is $\kappa$-invariant on the type IIA background [17][7] for $s = +1$, while it is also $\kappa$-invariant on the type IIA* background for $s = -1$. In other words, our Green-Schwarz action $I$ in (4.1) has the $\kappa$-invariance both on type IIA and type IIA* backgrounds, and in particular, the form of the action itself does not depend on the parameter $s$.

---
[11]See page 63 in [15], and page 172 in [16].

The fact that the same action (4.1) has consistent $\kappa$-symmetry both on the type IIA and type IIA* backgrounds is very suggestive that these two supergravity backgrounds are just different manifestations of a more fundamental theory connected by some duality, like the combination of T- and R-dualities, as indicated in [1].

## 5. Generalization of 'Star' Supersymmetry Algebras

As some readers may have already noticed, we may generalize our result to lower dimensions for extended supergravity theories.[12] Consider an unconventional $N$-Extended supergravity algebra in a given space-time dimensions $D$ with the ordinary Lorentzian signature[13] is

$$\{Q_{\underline{\alpha}}{}^i, Q_{\underline{\beta}}{}^j\} = \left[\, \varepsilon^{ij}_{(1)} (\gamma^a)_{\underline{\alpha\beta}} + \varepsilon^{ij}_{(2)} (\gamma^5 \gamma^a)_{\underline{\alpha\beta}} \,\right] P_a \ , \tag{5.1}$$

where the $N \times N$ matrices $\varepsilon_{(1)}$, $\varepsilon_{(2)}$ are diagonal and such that $P_a$ can be expressed as an appropriate contraction of the anti-commutator with some constant matrices. This is equivalent to a superspace wherein the spinor-spinor-vector torsion tensors reads

$$T_{\underline{\alpha\beta}}{}^a = i \left[\, \varepsilon^{ij}_{(1)} (\gamma^a)_{\underline{\alpha\beta}} + \varepsilon^{ij}_{(2)} (\gamma^5 \gamma^a)_{\underline{\alpha\beta}} \,\right] \ . \tag{5.2}$$

This clearly covers the standard and 'star' version of 10D type IIA supergravity theories. The existence of the 'star' theories informs us that we can generalize the $\varepsilon$-matrices to have an indefinite signature $(p, q)$, where $p$ (or $q$) is the number of the $+$'s (or $-$'s), so that $p + q = N$.

If such a superspace supergravity theory is to provide a background for a Green-Schwarz action, it is necessary that some axion field strength occurs. The most general form of this consistent with the result in (5.2) is

$$G_{\underline{\alpha\beta} c} = i \left[\, \varepsilon^{ij}_{(3)} (\gamma^a)_{\underline{\alpha\beta}} + \varepsilon^{ij}_{(4)} (\gamma^5 \gamma^a)_{\underline{\alpha\beta}} \,\right] \ . \tag{5.3}$$

Now we can write a Green-Schwarz action and ask the question of how many inequivalent ways there are to determine the various $\varepsilon$-matrices such that this action admits a $\kappa$-symmetry, complying with the superspace Bianchi identities. The eigenvalues of the four $\varepsilon$-matrices define the equivalence classes of the Green-Schwarz action. The 'canonical' representatives of these classes may be defined by re-scaling these eigenvalue so that they all take on only the values $\pm 1$ or 0.

---

[12]Similar idea has been also given in [1].

[13]It is also easy to extend such model to spaces with arbitrary signatures.

The solution to the question above contains the 'star' theories whenever appropriate eigenvalues of $\varepsilon_{(1)}$ and $\varepsilon_{(2)}$ are negative, as well the unitary theories described in [15] and [16] and some generalization of these two classes of models. So there are clearly classes of 'star' models associated with the equivalences classes of $\kappa$-invariant Green-Schwarz models. Like our result in the last section, only the torsions and axion supertensors determine the $\kappa$-symmetry of the standard GS action.

In lower dimensions, the situation is obviously richer as compare to the 10D theories. In fact, it has been known for a long time [15] and [16], that in 4D there exist Green-Schwarz models with $\varepsilon_{(1)}^{ij} = \delta^{ij}$, $\varepsilon_{(2)}^{ij} = \varepsilon_{(4)}^{ij} = 0$ but with $\varepsilon_{(3)}^{ij} \neq \delta^{ij}$. These are also members of the theories defined by (5.2) and (5.3). Unlike the 'star' theories in [1], the theories described by [15] and [16] define *unitary* field theories. Some of these unitary theories have been interpreted as closed GS strings constructed from left-handed open GS strings with $p$-supersymmetries and right-handed open GS strings with $q$-supersymmetries [18].

## 6. Concluding Remarks

In this paper, we have presented a systematic formulation for type IIA$^*$ theory, namely starting with the component invariant action with transformation rules, the corresponding superspace formulation has been established. We have also developed a compact superspace constraint notation presenting two sets of $\beta$FFC for type IIA and type IIA$^*$ theories in terms of just one signature parameter $s = \pm 1$ interpolating these two distinct systems. Based on such superspace backgrounds, we have also given the Green-Schwarz superstring action, that is consistent with such backgrounds. Interestingly, we have found that the conventional Green-Schwarz action for the type IIA background takes exactly the same form as that for the type IIA$^*$ background, suggesting some fundamental duality interpolating these two theories, such as the combination of the T- and R-dualities [1].

At first glance, the presence of kinetic terms for the 10D matter gravitino multiplet possessing the 'wrong' sign seems a serious drawback of type IIA$^*$ theory, due to the breaking of unitarity, positive definite energy, and causality, *etc*. However, we can remind ourselves that there have been previous interesting discussions of this type. Take for example, the background for the $N = 2$ superstring [7]. This background describes $N = 2$ self-dual supergravity in the Atiyah-Ward space-time in 4D with the signature $(+, +, -, -)$ [19]. Thus for such non-conventional superstring/supergravity theories, past experience showed relations to integrable systems in lower-dimensions, and such self-dual supergravity plays an important role. It is also important to notice that the pure 10D, N = 1 supergravity subsector of the type IIA$^*$ theory has unitarity, positive definite energy, and causality.

Although we have not performed a similar detailed analysis for type IIB$^*$ theory, it is clear that we can flip some signs in $T_{\underline{\alpha\beta}}{}^c$. Or to be more specific, take eq. (3.1) in [9] but instead of $T_{\alpha\overline{\beta}}{}^c$ non-zero while $T_{\alpha\beta}{}^c = T_{\overline{\alpha}\overline{\beta}}{}^c = 0$, we now choose the alternative choice, namely, the former to be zero, while the latter two to have opposite sings. Even though we do not give the result here, we can show that the signs for the kinetic terms of the RR fields in the gravitational equation have the 'wrong' signs.

Once we have understood that the maximal supergravity theory in 10D can be decomposed into $N = 1$ submultiplets, where the matter action is assigned with the 'wrong' sign, we can expect the similar mechanism in any other maximal as well as non-maximal supergravity theories in any lower dimensions, such as the 4D, $N \leq 8$ supergravity.

Even though our first presentations of lagrangians and superspace formulations are rather 'routine', we still have found something unconventional, like the fact that a single signature parameter $s = \pm 1$ can interpolate two systems of type IIA and type IIA$^*$ supergravity, and moreover, the same Green-Schwarz action has $\kappa$-symmetries consistent with such supergravity backgrounds.

We believe that our explicit result here will initiate further developments in these supergravity/superstring/supermembrane or D-brane physics in the near future.

## Acknowledgment

We are grateful to C. Pope and W. Siegel for helpful discussions.12